# A New Approach for Quality Management in Pervasive Computing Environments

ALTI Adel[1], ROOSE Phillipe[2]

[1] Computer Science Department, Science Faculty
Ferhat Abbas University, Sétif B.P. 19000, Algeria

[2] LIUPPA / IUT Bayonne
2 Allée du Parc Montaury 64600 Anglet – France

**Abstract**

This paper provides an extension of MDA called Context-aware Quality Model Driven Architecture (CQ-MDA) which can be used for quality control in pervasive computing environments. The proposed CQ-MDA approach based on ContextualArchRQMM (Contextual ARCHitecture Quality Requirement MetaModel), being an extension to the MDA, allows for considering quality and resources-awareness while conducting the design process. The contributions of this paper are a meta-model for architecture quality control of context-aware applications and a model driven approach to separate architecture concerns from context and quality concerns and to configure reconfigurable software architectures of distributed systems. To demonstrate the utility of our approach, we use a videoconference system.

## 1. Introduction

Model Driven Approach (MDA) [5] has been proposed by the OMG (Object management Group). The basic models of MDA are entities able to unify and support the development of computer systems by providing interoperability and portability. MDA approach does not address how to consider non-functional demands, i.e. how to represent and transform them.

An application for heterogeneous mobile embedded and limited (low bandwidth, power consumption, etc.) device has to firstly prevent interaction and mobility limitation. The heterogeneity of components regarding embedded sensors, CPU power, communication mechanisms (GPRS, WIFI, Bluetooth, ZigBee, etc.), speed of transmission as well as the media variety (sound, video, text and image) requires taking into account adaptation to an abstract level in order to avoid the ad hoc solutions which are not reusable and/or generalized. This is due to the following points:

- The separation of concerns met in software architecture is the separation of communications supported by first class connector from the business logic supported by components. However, communication is not the unique non-functional concern found in software design. Data adaptation, context-awareness, resource-awareness and QoS are other non-functional concerns which cut across component's business logic. Introducing in software architecture will make design of complex software an easier task and will yield clear and lucid specification.

- Few ADLs are able to define new connectors' types that ensure the non-functional concerns of the components (*security, communication, conversion, etc.*).

- Few ADLs support the elaboration of quality model explicitly and facilitate the system architecture quality control with the continuous evolution of its context.

In this paper, we present an extended Model Driven Architecture which includes support for software architecture quality control and resources requirements changes, in the framework of CQ-MDA (Context-aware Quality Model Driven Architecture). Some other works concentrate only on quality system architecture or context-aware system architecture [8, 9]. Our approach focuses on separation of two concerns: the architecture and the implementation contexts. This enables us to support them with the elaboration of quality model explicitly and to facilitate the system architecture quality control with the continuous evolution of its context. To cope with a serious gap in styles quality control, we have previously introduced the *ArchRQMM* (ARCHitecture Requirement Quality MetaModel) [3]. One of the strengths of *ArchRQMM* relies in its ability to separate architecture concerns from requirement and quality concerns and to automatically perform formal architecture quality analysis at architecture stage using OCL [12]. However, our metamodel does not support the definition of a context-awareness and a resource-awareness metamodel.

We begin this paper by introducing ArchRQMM metamodel. Section 3 proposes the main element of CQ-MDA approach, i.e. ContextualArchRQMM metamodel which it is an ArchRQMM extension used as support for context model description and quality model definition. Section 4 describes the CQ-MDA itself. Section 5 shows an example of applying CQ-MDA for VideoConference system development [15]. Section 6 summarizes related works. Section 7 concludes this article and presents some future works.

## 2. An Overview of ArchRQMM (Architecture Requirement Quality Metamodel)

*ArchRQMM* metamodel enables architectural styles quality evaluation and selection at the architecture design step and ensures formal verification of the properties' quality of architectures on modelling styles. The metamodel was described in details in [3, 4]. It was developed according to ISO/IEC 9126 standard [7]. *ArchRQMM* is based on a set of meta-classes for the common concepts of architectures descriptions languages (ADLs) and a set of quality characteristics based on a standard ISO quality model [10] which can be investigated and evaluated in the architecture level (maintainability, reusability, efficiency, etc.) . Fig. 1 presents a MOF metamodel of the *ArchRQMM*. One of the strengths of *ArchRQMM* relies in its ability to separate architecture concerns from requirement and quality concerns and to automatically perform formal architecture quality analysis at architecture stage using OCL [12]. The focus of rigorous architecture quality analysis is to prevent the non-required affections before the early phases of system development. The use of *ArchRQMM* metamodel offers number of advantages compared to other related works using UML profiling mechanisms like MARTE [18] including: 1) – architectures, requirements and quality models are explicitly represented, 2) – a formal support to prove the quality properties of architectural styles at the architecture level using OCL[12], 3)- support for model non-functional aspects of software architecture through architecture properties and measurable standards [7,4] , and 4) – automatic evaluation and selection of styles that best meet architects' needs using *QualiStyle* tool [4].

## 3. ContextualArchRQMM Metamodel

### 3.1 Objectives and Motivations

The main idea of this proposal is to take into consideration the non-functional concerns (*adaptation service, communication protocol, security, QoS, etc.*) of the components by connectors at the software architecture level. In our approach, the two types of preoccupations are ensured respectively by the components and the connectors. Thus, the connectors ensure the communication and the connection of components that realize the functional part (*business logic components*). Their execution within adequate configurations also requires taking into account of the non-functional aspects.

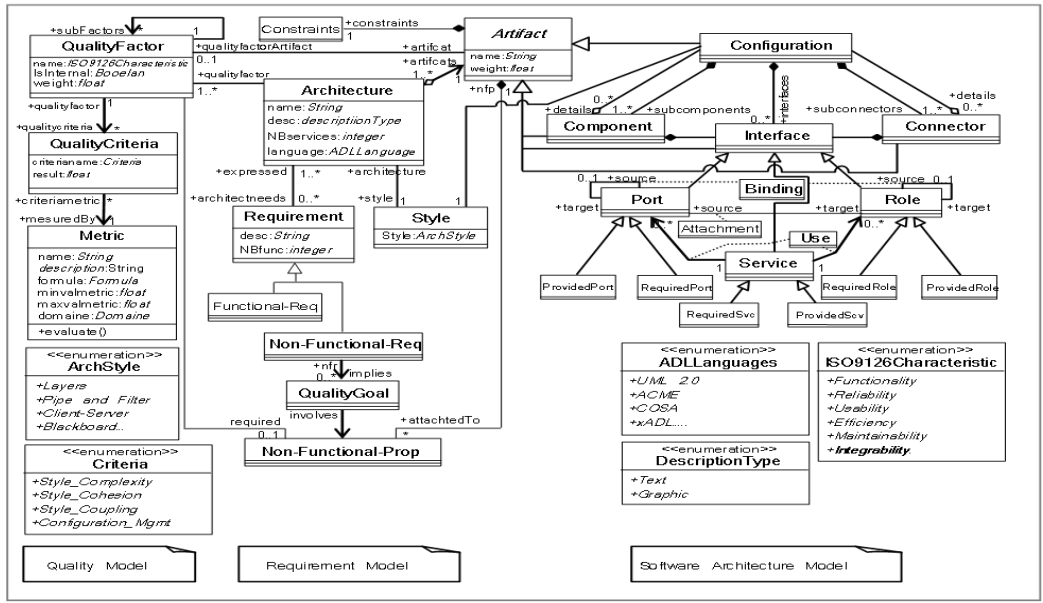

Fig. 1  A MOF Metamodel of ArchRQMM.

## 3.2 Context-awareness Metamodel

We extend our software architecture metamodel, with a context metamodel (Fig. 2). The goal is to represent context information of system architecture at model level. Context is any information that can be collected from artefact needs, resources capacities and user preferences [20]. *ContextualArchRQMM* uses these informations to perform a software architecture quality evaluation and selection in software development process. We have identified two types of context, i.e., required context (user preferences, artifacts needs) and provided context that encompasses the properties of the execution environment of an application. Context elements are realized through *Context* class, are expressed as QoS properties of the contextual artifacts (*Non-Functional-Prop* class).

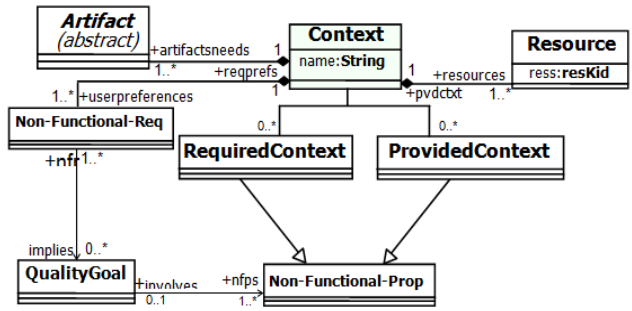

Fig. 2 The context metamodel of ContextualArchRQMM

## 3.3 Resource-awareness Metamodel

Fig. 3 depicts a resource-awareness metamodel. The hardware components are mobile devices (Class *Device*) like PDAs, PC Portables or smart phone, are constrained in their resources (memory size, CPU power, bandwith, battery, etc) and act as execution environment for architectural artefact (Class *Artifact*). Network connections (Class *Node*) connect hardware components having a limited bandwidth. A resource-awareness about current usage of processing power, network bandwith, etc. is a prerequisite to guarantee a minimum quality of service.

## 3.4 Contextual Architectural Artifacts

For an efficient and clear specification of connection points, we have introduced more precise port according to their global roles in a component: the *DataPort,* the *ContextPort,* information available at run-time when the service is active. The *ServiceControlPort* is a standard dedicated port for controlling a service. It allows the service to be (re)started, updated, relocated, stopped and uninstalled.

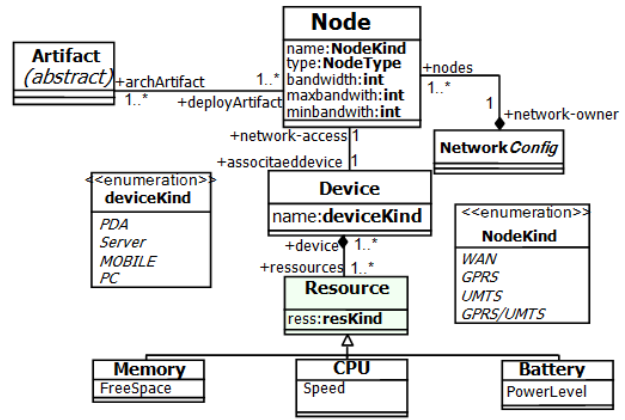

Fig. 3 The resource metamodel of ContextualArchRQMM

The *QoSNotificationPort* is responsible for sending QoS information to execution platform in order to decide if a service reconfiguration is needed. As software architecture descriptions rely on a *connector* to express interactions between components, an equivalent abstraction must be used to express a contextual and a heterogeneous interaction (i.e. various interactions paradigms). We extend an architectural connector with a contextual concern in a heterogeneous interaction (Fig. 4). Three auto-adaptative mechanisms are distinguished: communication (i.e. clarify the connection between various components regarding the communications paradigms), service adaptation (i.e. adding, suppression and substitution of adaptation services), and QoS adaptation (selecting parameters of service to provide adequate quality to component needs at runtime). The business logic component is adapted explicitly and automatically by a *contextual connector*. This means that context ports of *business logic* components instances, related to the context managed by a contextual connector, are all connected to that contextual connector. The *data role* may be connected to the *data port* of a component (provided or required) and the *contextual role* may be connected to the *contextual port* of a component. The distinction between a data and context roles (and also between a data and context ports) addresses the constraint typically imposed by many ADLs about the clear separation between functional and non-functional aspects. This ensures a quality of the components assembly by inserting a contextual connectors, as well as management of adaptation service quality.

## 3.5 Metamodel for Dynamic Reconfiguration

Dynamic reconfiguration is defined by transitions between configuration families (Fig.5.). Our metamodel proposes to define *configuration family* to capture a non-predefined number of configurations having close adaptation services.

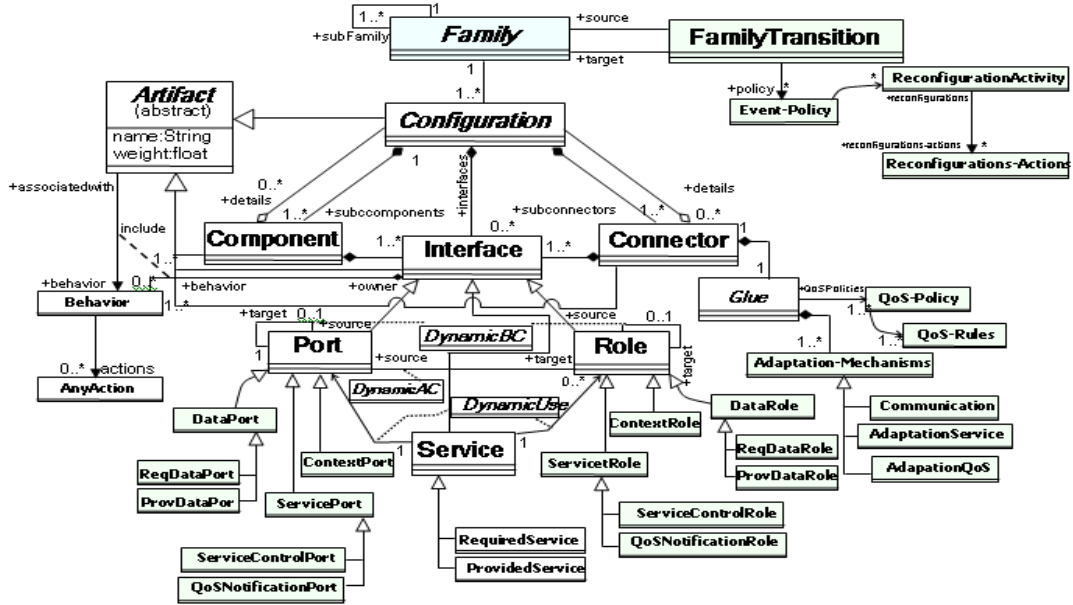

Fig. 4 Contextual architectural artifacts in ContextualArchRQMM

For each family, a specific set of adaptation services defined. For example at *image family* which includes connectors offering services of the same nature (i.e. image adaptation services) but only differs by their adaptability to the context.

A transition allows switching the system from the source configuration family to another new target configuration family. A transition can be triggered by different events, like changes in the environment, changes in the applications to be executed, or changes in the system operational conditions (e.g., a battery operated system detects a change in the battery status, or a component that becomes faulty). We can have a transition into the same configuration family; it is a transition between two configurations of the same family. For each transition, a reconfiguration activity presenting a set of reconfiguration actions is associated. It represents a set of actions switching from the current configuration to the target one. In our approach we have a non-predefined number of configurations, but we have statically predefined families. To answer to an adaptation task, on a mobile device system at the run-time, one needs to satisfy a new need related to a new execution context. The ideal solution is to install, update or remove an adaptation service at the connector's configuration. This contribution of reconfiguration is similar to other work described in a paper [11] but our work concentrates on connector reconfiguration and insisted on the separation of the two concerns: software architecture model and context model.

Four possible adaptations in *ContextualArchRQMM* are: parametric adaptations (i.e. an update parameter value command is sent along with the name and the new value of the parameter to the command queue of the connector), services adaptations (i.e. call to another available service provider by *composing* and/or *decomposing* of services using the *DynamicUse* concept), sub-family (re) assembly: (i.e. attach/detach several subfamilies into a family), move and re-routing**:** (i.e. we use the routing service to lookup another relay to deploy the desired service).

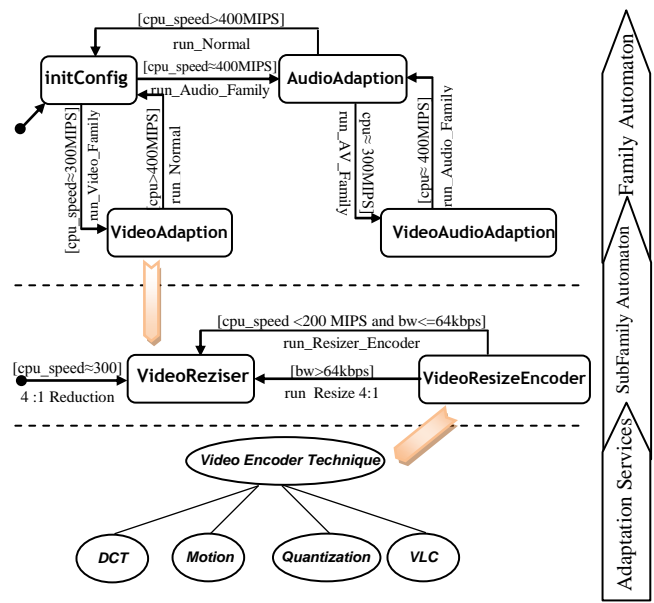

Fig. 5 Autmoaton hirerachy in the adaptation connector

# 4. Context-aware Quality – Model Driven Architecture (CQ-MDA)

The general structure of Context-aware Quality – Model Driven Architecture (CQ-MDA) is presented in Fig. 6. We consider the full software development cycle within MDA, i.e. from formulation of needs up to the code generation. The proposed structure consists in five levels representing CIM, PIM, Contextual Platform Independent Model (CPIM), Contextual Platform Specific Model (CPSM), and code. Each level is decomposed into three parts: the left part represents architectural artifacts and context concepts; the right part represents quality model and measurements done for these artifacts while the center part represents requirements.

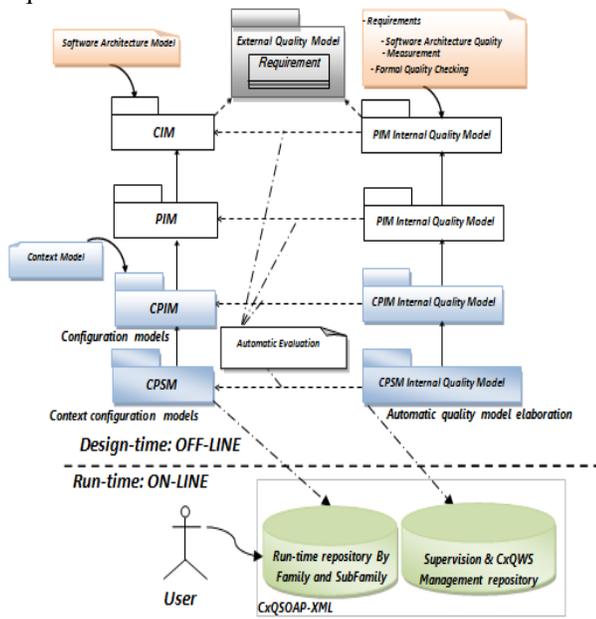

Fig. 6 Context-aware Quality Driven Model Architecture

## 4.1 Architecture Quality Control at the Design-Time

Architecture quality should be controlled at each steps of the design. *External requirement*s of the system are transformed into *internal ones* for the architecture and its components. *Internal requirements* are needed for assessing designed architecture models. So, particular internal models, being instances of *ContextualArchRQMM* metamodel, are used to assess particular models of CQ-MDA. The software architecture quality model is produced by measurement done for each architectural artefact for a given factor in the context of associated requirement, for a given criteria with associated metric. Two ways of using our meta-model are possible:

- The first one assumes that the software architecture quality metamodel is used for evaluating an architecture model. The architecture model is tested and validated with the semantic constraints defined by the metamodel. If the verified architecture model gets bad marks then the design process can be stopped or it can go back to the previous stage either to change requirements or to elaborate a different (better) architectural model.

- The second one, using software architecture quality metamodel considers the case when the metamodel is used for selecting the best architectural model from different choices. In this case the values of a metric are used to classify the models. A metric formula gives a note for the architecture model. The values of the metric function are used to classify the models and to choose the suitable one and we select a first model if we have the same value. After that, the selected architectural model is evaluated by the OCL constraints to remove any quality semantic violation.

## 4.2 Architecture Adaptation at the Run-Time

We can say that two configurations provide a close service if and only if their marks of the architecture quality criteria (i.e. context-independent) and contextual architectural quality criteria (which are related to run-time context) are close. Because context-independent quality criterion variation is more perceptible by users, platform will begin its research with the evaluation of the configurations having the same mark of context-independent quality criterion as the current configuration. In response to events notifying about changes in the environment (less bandwith, less available memory…), or in the running application (overflow/underflow of the buffer, increased transmission time…), the *Adaptation Manager* will be notified by set of probes which constitute the monitoring framework, update configurations and annotate the events to these configurations. Our platform use configurations families and subfamilies described in XML format from a preliminary analysis of the application (i.e. at the design step) in terms of QoS and update it in real time.

For an efficient and better implementation of self-management process (Fig. 7), we have used *"poisson"* simulation and formal methods (OCL) to assess the degradation of quality attributes due to movement of devices and employ runtime adaptation to mitigate such problems.

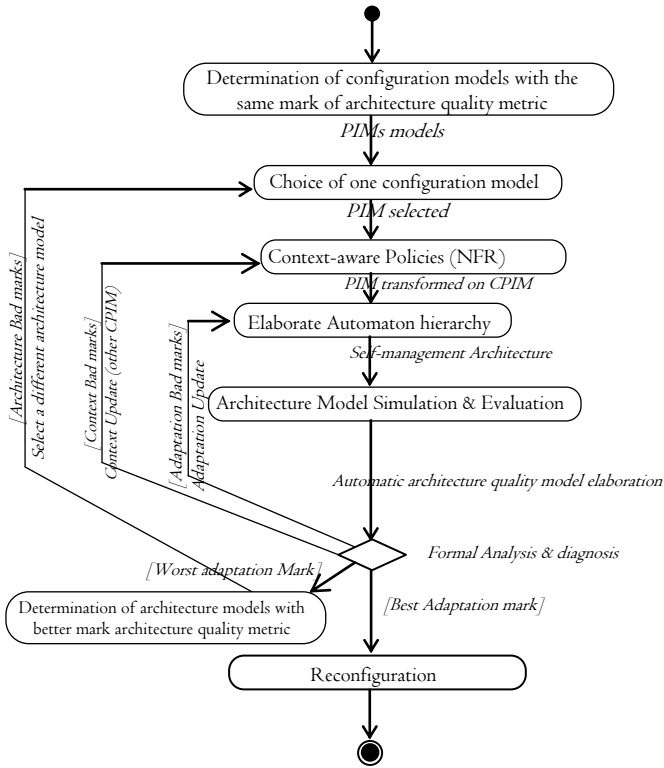

Fig. 7 Process reconfiguration model of CQ-MDA

Our process started with the evaluation of the configurations having the same mark of structural architecture criterion *(coupling, cohesion, structural complexity…,)* as the current configuration. That will only modify the mark of the adaptative criterion (*response time, adaptation effort…*). As soon as a reconfiguration event is received, the *Quality-Manager* search for a better configuration model to using successively by analyzing finite sets of configurations *having the same mark of structural architecture quality metric and differ only by their adaptability cost to the context.* Firstly, the platform will be able to restrict the scope of the search into the range of configurations, which differ from the current configuration only by the adaptation service (or component) at the origin of the reconfiguration event. But when this approach does not give any solution, we face the issue of the deployment of a sub-family or a family. The *Adaptation Manager* receives the new selected configuration model and starts-up the reconfiguration.

## 5. Case Study: Video Conference System

A case study given below is intended to show applicability of CQ-MDA both for evaluation, for selection and for reconfiguration of the best architectural model from some alternatives.

A case study deals with *VideoConference* System [15]. *VideoConference* has the following optional services:
- Audio Encoder: (de-)compressing the audio stream.
- Video Encoder: (de-)compressing the video stream.
- Audio Filter: components for changing the frame size.
- Video Filter: reducing the video frame rate.

The following user preferences are considered:
- Recording, reviewing user' video and creating respective reports.
- Video should be delivered in quality and in period no longer than one minute from their request.

According to *ContextualArchRQMM*, all these requirements should be associated with a respective architecture quality model with selected quality factors. In our example, for illustration, only non-functional requirements are taken into account. It is proposed to use the efficiency factor with time-behavior sub-factor [4]. On the CIM level some internal requirements may be specified additionally to external ones. We propose "an easy maintenance of software architecture model: internal requirement" as we consider it to be important factor from architect point of view. This additional requirement can be expressed more precisely as "low complexity, high cohesion and low coupling these requirements are the main facts to take into account for achieving easy maintainability architecture (subfactors of the maintainability factor [4])." The time behaviour sub-factor for software architecture model artefact cannot be evaluated at CIM level (as the software architecture is not defined yet) and should be forwarded to the next level i.e. PIM level. Therefore the CQ-MDA approach will be shown in details using the transformation of the PIM model with respective internal quality model into CPIM model with its internal quality model and the CPIM model with respective internal quality model into CPSM model with its internal quality model.

### 5.1 PIM Level – Quality Control at the Design-Time

PIM model is the starting point for the considered transformation. Several architectural models can be used to design a given system. For the *VideoConference* system, the model is designed with *PipesAndFilters* style as shown in Figure 8. At PIM level we have also formally defined set of architectural artifacts that are traced from CIM model.

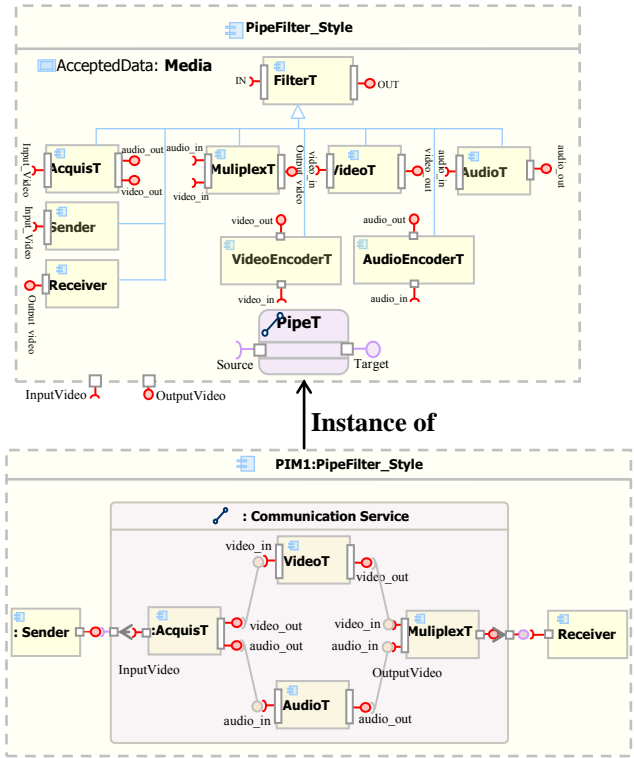

Fig. 8 PIM software architecture model

Internal quality model on this level is traced from the upper quality level model. So, we have to consider the factors from CIM level, i.e. efficiency factor with time-behaviour sub-factor and maintainability factor with modularity, analyzability sub-factors. The first factor is efficiency with sub-factor Time-behavior cannot be evaluated at this level as we have not found accepted metrics for evaluation of the PIM model. This factor must be still forwarded for evaluation to the next modeling level. The second factor is maintainability with modularity and analyzability sub-factors [4]. The first sub-factor, modularity, depends on the configuration, component and connector modularity. If the system has been divided correctly to suitable modular, the software system can be analyzed more easily. At the architecture level, this factor can be measured with criteria, named *coupling* and *cohesion*. In [4] these two metrics are proposed for measuring architecture modularity. We used these metrics in our model. We have evaluated each kind of models with similar measurements of the whole architecture of the basic metrics (i.e. coupling, cohesion and complexity).

The evaluation results are given in Tab. 1 using a prototype implemented written in Java called *QualiStyle* [4]. The architecture model should be tested and validated with the semantic constraints defined by the meta-model. If the verified architecture model gets bad marks then the design process can be stopped or it returned to the previous stage (i.e. CIM) either to change requirements or to elaborate a different (better) architectural model. High cohesion, low coupling and low complexity are the main facts to take into account for making a design understandable, maintainable, and of higher quality. All these basic metrics are in *[0, 1]*. The higher cohesion's value (resp. lower complexity's value) is the better for architecture quality. As for the architecture model from Table 1 the values of coupling is equal 0.482 and a threshold of coupling is equal 0.66, the value of cohesion is equal 0.341 and a threshold of coupling is equal 0.5 and the value of complexity is equal 0.362 and a threshold of complexity is equal 1, the architectural model provides an acceptable maintainability (a high level of cohesion, a low level of coupling, a low level of complexity). This architectural model is accepted for further transformation. This result is practically significant as well related to maintainability effort, e.g. low level of coupling, dependencies among all architectural artifacts are loss, high number of reused artifacts (i.e. number of Pipe connector instances, m = 4).

Table 1: PIM evaluation results.

| *PIM* | *Coupling* | *Cohesion* | *Complexity* |
|---|---|---|---|
| Pipe-Filter | 0.482 | 0.341 | 0.362 |

### 5.2 CPIM Level–Quality Control at the Design-Time

PIM software architecture model may be transformed, manually or automatically, into different CPIM models. The software architecture model from Fig. 8 is transformed into five CPIMs models (Fig. 9) and the total resource requirements are given in Table 2. Fig. 10 depicts our automaton for the video adaptation family.

At this level analyzability, time-behavior sub-factors taken from upper level are evaluated (it is worth to mention – different metrics can be used for this purpose). The evaluation results should be helpful in choosing the best CPIM model for further transformation.

Table 2: resources requirements

| *Component* | *User preferences* | *CPU speed* | *Bandwith* |
|---|---|---|---|
| RateAudioT | - | ≈ 100 MIPS | 4:1 Reduction |
| ResizeVideoT | - | ≈ 400 MIPS | 2:1 Reduction |
| AudioEncoderT | High Quality<br>Medium Quality<br>Low Quality | ≈ 300 MIPS | 64 kbps<br>32 kbps<br>8 kbps |
| VideoEncoderT | High Quality<br>Medium Quality<br>Low Quality | ≈200 MIPS | 10:1 Reduction<br>20:1 Reduction<br>30:1 Reduction |

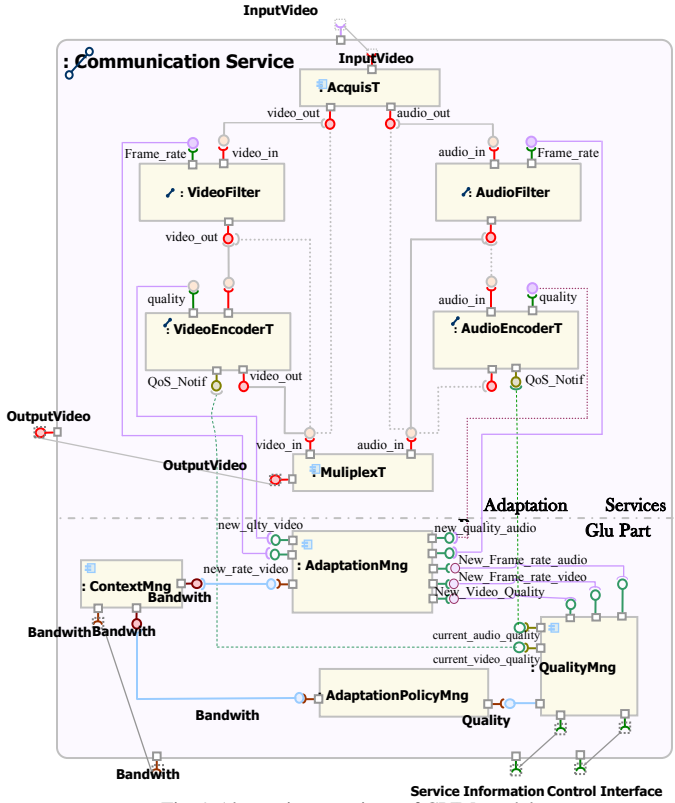

Fig. 9 Alternatives versions of CPIM models

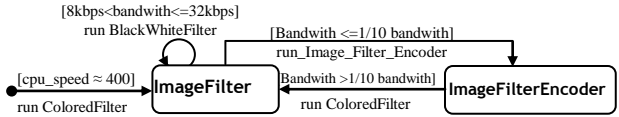

Fig.10 Video adaptation automaton

For time-behavior, three metrics proposed in [7], one of them is selected and adapted in our case. The estimated Time Behavior Metric (*TBM*) for a set *A* of artifacts of a given configuration performed with a given time in a certain context calculated as the weighted sum of $TB_a$ metric counted for every artefact instance *"a"*:

$$TBM^{Benefit}_{Memory_{size},CPU_{speed},Network_{bw}}(config) = \sum_{a \in A} w_a * TB_a \quad (1)$$

Apart from the evaluation of time behavior sub-factor we evaluate the analyzability sub-factor to select the best CPIM model. In [16] two metrics were proposed for the dynamic adaptivity at the architectural level, but only one, *MaAC* (Minimum architectural Adaptive Cost) was used and validated for analysability assessment in our example. According to the choice made of the sub-factors of quality and their measurement, we define the *Quality* function which measures the quality of a given configuration:

$$Quality(config) = \frac{TBM^{Benefit}_{Memory_{size},CPU_{speed},Network_{bw}}(config)}{MaAC^{Cost}_{Memory_{size},CPU_{speed},Network_{bw}}(config)} \quad (2)$$

Table 3 shows the evaluation results, meaning that CPIM5 turns out to be the best. Differences can be seen in the adaptation cost of this CPIM and other CPIMs, which is due to the low adaptation effort compared to other CPIMs. This result is practically significant as well related to adaptation effort e.g. number of artifacts which should be added to make a system adaptive are very loss as consequence of self- management for environment evolution (i.e. *CPU usage, bandwith*) guided by the adaptation policies.

Table 3: CPIMs evaluation results

| *Adaptable and optional services* | *TBM (ms)* | *MaAC (artifact nb)* |
|---|---|---|
| Video Resize, High Quality Video Encoder/Decoder | 200 ~ 400 | 0 ~ 16 |
| Video Resize, Medium Quality Video Encoder/Decoder | 200 ~ 330 | 0 ~ 16 |
| Video Resize, Low Quality Video Encoder/Decoder | 350 ~ 500 | 0 ~ 8 |
| Video Resize, Audio Encoder/Decoder | 470 ~ 800 | 0 ~ 8 |
| All Adaptable Services | 420 ~ 930 | 0 |

5.3 Architecture Adaptation at the Run-Time

Participants to the video conference are interested for service quality in the face of device heterogeneity. We distinguish two Participants' families: speaker and auditor. The service quality requirement can be satisfied by using our context quality management strategy. The goal for a given mobile device is to achieve qualities and allocate resources to result in the best configuration such that the system quality is maximized subject to device resource constraints, user preferences constraints. The platform is capable of adding/removing/updating/moving services at the execution time. The important task of our platform is to perform the dynamic changes at the run-time and, more precisely, with minimum length of time and decision making. It is necessary to have a mechanism for media flow measurement which will detect when the application must be reconfigured for reasons of lower available bandwidth. In addition, it is necessary to know when the bandwidth is sufficient to switch to another configuration. So, we propose to use our context quality management. We can see the different adaptations in the following scenarios:

**Scenario # 1**. The application is first of all deployed in a favorable context, where neither the stations nor the network are saturated. Initially, the context is sufficient to provide both video and audio. If we receive a video stream packaged with *RealVideo* in a 120 x120 window at 10 frames / second with phone audio quality, rate of 56Kbits is sufficient.

**Scenario # 2**. The supervisor has noticed a problem of bandwith, and thinks that the bandwith will not hold until the end of the video. To detect a decrease in media throughput, the *Adaptation Manager* receives two events of the buffering connector corresponding to *overflow/underflow* of the buffer size (i.e. 20% - 80% of the buffer size) from the supervisor. When an event of underflow is received, it indicates a problem of the video transmission (loss of information transmission, increased transmission time). Since, an overflow event implies that the current bandwith is not sufficient. To alleviate too many changes (i.e. minimum reconfiguration cost) in the current configuration, the application can switch to the ideal configuration if the video stream of data can be supported for long enough time (*depending on the size of the buffer*). The ideal management on bandwith degradation is to follow a minor change by the replacement of a service connector (*Video Encoder/Decoder with High quality*) by another connector service (*Video Encoder/Decoder with Lower quality*).

**Scenario # 3**. In another scenario, due to movement of devices, the network throughput connecting the devices is very loss, making it difficult for communication service to interact with auditor. The platform looks for a new configuration to use, starting by looking for a new relay allowing the moving a video resize connector to a suitable device.

## 6. Related Works and Discussion

The first related area of research are ADLs that have been proposed for representing dynamic architectures including: ACME [14], π-ADL [6], C3 [2] and AADL [1]. However, except for ACME, most ADLs do not support the concept of evaluation function. In addition, most of them are not contextual defined. MARTE [17] does not treat the problem of heterogeneity by a meta-model which verifies the adequacy of service regarding its context and research of the adaptation strategy [19]. Π-ADL [6] is a formal architecture description language based on the π-calculus. It does not support contextual connectors and not integrate quality metrics. Recently, Garlan and al. [14] extended ACME ADL in order to support evaluation function in evolution styles and their multiple decision forms. However, this work does not consider exploiting contextual connectors in heterogeneous environment where entities of different nature collaborate: software and hardware components. The second related area of research are some works involving quality in MDA approach, like QADA (Quality-driven Architecture Design and Quality Analysis) [8] – a methodology targeted at the development of service architectures. Other works involving Context in MDA approach, e.g. Context-aware Model Driven Architecture Model Transformation [13] – a methodology targeted at the development of context-aware applications and other networked systems. These works concentrate only on quality system architecture or context-aware system architecture, while CQ-MDA insisted on the separation of the two concerns: software architecture model and context model.

## 7. Conclusion and Future Works

This paper proposed *ContextualArchRQMM* metamodel centred on the concept of contextual connector, which take advantage of traditional architectural connectors and provides a lightweight support for the definition of some composition facilities such as heterogeneous interfaces at the connector level. The paper proposed also CQ-MDA approach based on ContextualArchRQMM, being an extension to the MDA, allows for considering quality and resources-awareness while conducting the design process. The main idea of presented extension consists of three abstractions levels: PIM, CPIM and CPSM. At the PIM level, a model is decomposed on two interrelated models: software architecture artifacts, which reflect functional requirements and quality model. At the CPIM level a simultaneous transformation of these two models with contextual information details are elaborated and then refined to a specific platform at the CPSM level. Such a procedure ensures that the transformation decisions should be based on the quality assessment of the created models. At design-time, our approach is used to assess the quality attributes of the system's architectures. At run-time, the framework copes with the challenges posed by the highly dynamic nature of mobile systems through continuous monitoring and calculation of the most suitable architecture. If a better architecture is found, the framework adapts at run-time the software, potentially via connector adaptation and mobility. We presented an illustrative example to show the applicability of the proposed CQ-MDA approach. The results of the experiments (based on the example of *VideoConference* with four CPIMs) are encouraging. The experiment shows that our approach outperforms two abstractions level in terms of some quality metrics such as adaptation ratio and time response. In the future, we will consider moving our approach to a real execution platform to validate its feasibility.

**Acknowledgments**

We would like to thank our colleagues and students for testing our simulation tool and specially *Hamza Reffad* and *Bekakchi Youcef*. The authors appreciate the in-depth comments given by the anonymous reviewers to improve this work.

**References**

[1] B. Berthomieu1, J.P. Bodeveix, C. Chaudet, F. Vernadat, "Formal Verification of AADL Specifications in the Topcased Environment," 14th Ada-Europe International Conference, 2009, pp. 207 – 221.

[2] A. Amirat and M. Oussalah, "First-Class Connectors to Support Systematic Construction of Hierarchical Software Architecture," Journal of Object Technology, Vol. 8. N°.7, 2009, pp. 107-130.

[3] A. Alti, A. Boukerram and A. Smeda, "Architectural Styles Quality Evaluation and Selection," 9th International Conference NOTERE'09, Montréal (Canada), 2009.

[4] A. Alti, A. Smeda, "Architectural Styles Quality Evaluation and Selection," Proceeding of 4th International Conference on Software and Technologies (ICSOFT'2009), Barcelona (Spain), 2009, pp. 74 - 82.

[5] J. Miller, J. Mujerki, editors. "MDA Guide, Version 1.0. OMG Technical Report,", http://www.omg.org/docs/ptc/03-05-01.pdf, 2003.

[6] F. Oquendo, "π-ADL: an architecture description language based en the higher order typed π-calculus for specifying dynamic and mobile software architecture," ACM Soft. Eng., vol. 29, n°. 4, 2004, pp. 1 - 13.

[7] ISO/IEC 9126-3. In Software Engineering – Product quality – Part 3: Internal metrics, ISO-IEC, 2003.

[8] QADA, http://virtual.vtt.fi/qada , 2007.

[9] P. Tarvainen, "Adaptability Evaluation at Software Architecture Level, " The Open Soft. Eng. J. vol. 2, Bentham Sc. Pub. Ltd., 2008, pp. 1-30.

[10] F. Losavio, L. Chirinos, N. Lévy, and A. Ramdane Cherif, "Quality characteristics for software architecture," JOT, 2(2), 2003, pp. 133-150.

[11] F. Kritchen, B. Hamid, B. Zalila and B. Coulette, "Designing Dynamic Reconfiguration for Distributed Real Time Embedded Systems," 10th International Conference NOTERE'2010, Tozeur (Tunisia) ,2010, pp. 249-254.

[12] OMG. UML OCL 2.0 Specification: Revised Final Adopted Specification. http://www.omg.org/docs/ptc/05-06-06.pdf, June 2005.

[13] S. Vale, S. Hammoudi, Context-aware Model Driven Development by Parameterized Transformation. MDISIS'2008, pp. 167–180.

[14] D. Garlan, J.M. Barnes, B. Schmerl, O. Celiku., "Evolution Styles: Foundations and Tool Support for Software Architecture Evolution," WICSA'09, 2009, pp. 16-25.

[15] S. Laplace, M. Dalmau, P. Roose, Prise en compte de la qualité de service dans la conception et l'exploitation d'applications réparties, In the Workshop GEDSIP@Inforsid 2009, Toulouse, 26 mai 2009.

[16] C. Raibulet, L. Masciadri, "Evaluation of Dynamic Adaptivity through Metrics: an Achievable Target?" WICSA'09, 2009, pp. 65-71.

[17] S. Gérard, D. Petriu and J. Medina. "MARTE: A New Standard for Modeling and Analysis of Real-Time and Embedded Systems", 19th Euromicro Conf. on Real-Time Systems (ECRTS 07), Pisa, Italy, 2007.

[18] OMG. A UML Profile for MARTE: Modeling and Analysis of Real-Time Embedded systems, June 2008, http://www.omg.org/docs/ptc /09-06-08.pdf , 2008.

[19] C. Marcel, R Michel, Christian M. "Autonomic Adaptation based on Service-Context Adequacy Determination". In ENTCS, p. 35-50, 2007.

[20] M. Dalmau, P. Roose, S. Laplace. "Context Aware Adaptable Applications - A global approach", Special Issue on Pervasive Computing Systems and Technologies - International Journal of Computer Science - IJCSI Vol. 1, Issue 1, 2009 - ISSN 1694-0784

**Adel Alti** obtained the Master degree from the University of Setif (UFAS), Algeria, in 1998. He is holding a Ph.D. degree in software engineering from UFAS university of Sétif, Algeria, 2011. Right now he is an associate professor at University of Sétif. He is a member of the research group LRSD. His area of interests includes automated software engineering, mapping multimedia concepts into UML, semantic integration of architectural description into MDA platforms, context-aware quality software architectures and automated service management, Context and QoS. During his work he has published number of publications concerning these subjects.

**Roose Philippe** is an associate professor at the LIUPPA/UPPA – FRANCE. He obtained his PhD degree in computer science from university of Bayonne, France, 2001. He head of the MOJITO and AEXIUM projects. His research interests are software architecture and platforms, pervasifs and ubiquitous computing, mobility, software components services, context and QoS, multi-parts profiles. He is the co-author of three books on software component technologies.